 \documentclass[12pt]{iopart}
\usepackage{iopams,amssymb}
\eqnobysec

\begin{document}
\jl{1}

\def\a{\alpha}
\def\b{\beta}
\def\d{\delta}
\def\e{\epsilon}
\def\g{\gamma}
\def\k{\kappa}
\def\l{\lambda}
\def\m{\mu}  
\def\o{\omega}
\def\t{\theta}
\def\s{\sigma}
\def\D{\Delta}
\def\L{\Lambda}


\def\beq{\begin{equation}}
\def\eeq{\end{equation}}
\def\bea{\begin{eqnarray}}
\def\eea{\end{eqnarray}}
\def\ba{\begin{array}}
\def\ea{\end{array}}
\def\no{\nonumber}
\def\le{\langle}
\def\re{\rangle}
\def\lt{\left}
\def\rt{\right}

\renewcommand{\theequation}{\arabic{equation}}

\title{Jordan-Wigner fermionization for the one-dimensional Bariev model of
three coupled XY chains}

\author{X.-W.~Guan$^{\dag}$, A.~Foerster$^{\dag}$,   
J. Links$^{\ddag}$ and H.-Q.~Zhou$^{\ddag}$}
\address{$^{\dag}$Instituto de F\'{\i}sica da UFRGS,
                     Av.\ Bento Gon\c{c}alves, 9500,\\
                     Porto Alegre, 91501-970, Brasil}
\address{$^{\ddag}$Centre for Mathematical Physics, 
School of Physical Sciences, \\ The University of Queensland, 4072, Australia}

\begin{abstract}
The Jordan-Wigner fermionization for the one-dimensional Bariev model
of three coupled XY chains is formulated. The Lax operator in terms of
fermion operators and the quantum R-matrix are presented
explicitly. Furthermore, the graded reflection equations and their
solutions are discussed.
\end{abstract}



\maketitle

It is well known that the Jordan-Wigner transformation is generally
used to convert spin models into fermion models, or vice versa, in 
condensed matter physics. Successfully applied to the one-dimensional
Hubbard model, the Jordan-Wigner transformation led to a complete
parametrization of the quantum R-matrix \cite{Sha,Wad1} in the
framework of the Quantum Inverse Scattering Method (QISM)
\cite{QISM1,QISM2}. However, the generalization of it to integrable
systems with an arbitrary number of internal degrees of freedom seems
to be rather infrequent because of the cumbersome calculations
involved. Although there is a scheme proposed by G\"{o}hmann and
Murakmi \cite{GM} to treat the fermionization of integrable lattice
systems, it only applies to a model where the Lax operator coincides
with the quantum R-matrix. Another generalization of the Jordan-Wigner
transformation was proposed recently by Batista and Ortiz \cite{JWbo}
to convert spin-$S$ operators of the $SU(2)$ algebra into fermion
operators. Although it reveals a new insight into the integrable
models of strongly correlated electrons, it can just be applied to
higher spin representation.  Nevertheless, some very interesting
models do not present this property, such as the 1D Hubbard model
\cite{Sha,Wad1}, the two-coupled XY chains \cite{Bar1,zhoutwo}, three
coupled XY chains \cite{zhou97}, etc.  In this communication, we
generalize the Jordan-Wigner transformation \cite{Wad1,zhou88,JW2} to
the one-dimensional Bariev model of three coupled XY chains
\cite{Bar1,Bar2,zhou97}, which possesses three internal degrees of
freedom. The model has finite magnetization at the ground state in 
zero external field and exhibits the existence of hole pairs of
Cooper type which are relevant to theories of superconductors
\cite{Bar2}. We convert the three coupled spin chains into a fermion
model of strongly correlated electrons. The graded Yang-Baxter
relation and the graded reflection equations, which guarantee the
integrability of model in the bulk and at the  boundaries, respectively,
are formulated in the framework of QISM. The integrable boundary
conditions for the fermion model, as well as the first conserved
current next to the Hamiltonian and the boundary interactions in terms of
fermion operators, are constructed.  Our results facilitate the
algebraic Bethe ansatz \cite{Mar} for the model with periodic and open
boundary conditions, which provides the spectrum of all
conserved charges,  essential in studying finite temperature
properties of the integrable models.

We begin by considering  a spin chain model defined by the following Hamiltonian
\begin{equation}
H = \sum ^{L}_{j=1} H _{j,j+1}\label{Ham-s}
\end{equation}
where $H_{j,j+1}$ denotes the  Hamiltonian density of
three $XY$ spin chains coupled to each other \cite{zhou97}
\begin{equation}
\fl
H_{j,j+1}  =  \sum _{\alpha=1}^3\;\left(\sigma^+_{j(\alpha)}\sigma^-_{j+1(\alpha)}+
\sigma^-_{j(\alpha)}\sigma^+_{j+1(\alpha)}\right)
\exp\left(\eta \sum _{\alpha^{'} \neq \alpha }
\sigma ^+_{j+\theta(\alpha^{'}-\alpha)(\alpha^{'})}
\sigma ^-_{j+\theta(\alpha^{'}-\alpha)(\alpha^{'})}\right),\label{Ham-sd}
\end{equation}
where $\sigma ^{\pm}_{j(\alpha)}= \frac {1}{2}(\sigma ^x_{j(\alpha)}
\pm i\sigma ^y_{j(\alpha)}) $, with $\sigma ^x_{j(\alpha)},
\sigma ^y_{j(\alpha)},\sigma ^z_{j(\alpha)}$ being the usual Pauli 
spin operators at site $j$  corresponding
to the $\alpha$-th ($\alpha=1,2,3$) $XY$ spin chain,
$\theta(\alpha^{'}-\alpha)$ is a step
function of $(\alpha^{'}-\alpha)$ and $\eta$ is a coupling constant.
As was shown in \cite{zhou97},
the Hamiltonian  commutes with  a one-parameter 
family of transfer matrices $\tau(u)$
of a two-dimensional lattice statistical mechanics model.
This transfer matrix is the trace of a monodromy
matrix $T(u)$, which is defined , as usual, by 
\begin{equation}
T(u)=L_{0N}(u)\cdots L_{01}(u) \label{TM}
\end{equation}
with $L_{0j}(u)$ of the form,
\begin{equation}
L_{0j}(u) = L^{(1)}_{0j}(u) L^{(2)}_{0j}(u) L^{(3)}_{0j}(u),\label{mono}
\end{equation}
where
\begin{eqnarray}
\fl L^{(\alpha)}_{0j}(u)& =& \frac{1}{2}(1+\sigma ^z_{j(\alpha)}
\sigma ^z_{0(\alpha)}) + \frac{1}{2} u (1-\sigma ^z_{j(\alpha)}
\sigma ^z_{0(\alpha)})
\exp\left(\eta \sum^{3}_{ \mbox{\scriptsize $\begin{array}{l}\alpha^{'}=1 \\
\alpha^{'} \neq \alpha \end{array}$}}
\sigma ^+_{0(\alpha^{'})}
\sigma ^-_{0(\alpha^{'})}\right)\nonumber\\
\fl & &+(\sigma^-_{j(\alpha)} \sigma^+_{0(\alpha)} + \sigma ^+_{j(\alpha)}
\sigma ^-_{0(\alpha)}) \sqrt {1+
\exp\left(2 \eta \sum^{3}_{\mbox{\scriptsize $\begin{array}{l}\alpha^{'}=1 \\ \alpha^{'} \neq \alpha\end{array}$ }}
\sigma ^+_{0(\alpha^{'})}\sigma ^-_{0(\alpha^{'})}\right) u ^2}. 
\end{eqnarray}
The explicit form of the corresponding R-matrix is given  in \cite{zhou97}.

Now let us introduce the following Jordan-Wigner transformation for a model with three degrees of freedom 
\begin{equation}
\left(\begin{array}{l}\sigma ^+_{j(\alpha)}\\ \sigma ^-_{j(\alpha)}\end{array}
\right) =
[V_{j(\alpha)}]^2\left(\begin{array}{l}c^{\dagger}_{j(\alpha )}\\
c_{j(\alpha )}\end{array}\right),\,\,\,\,
\sigma^z_{j(\alpha)}=2n_{j(\alpha)}-1,\label{J-W-T}
\end{equation}
where 
\begin{equation}
V_{j(\alpha)}=\left(\begin{array}{cc}v_{j(\alpha)}&0\\0&v^{-1}_{j(\alpha)}\end{array}\right)
\end{equation}
with
\begin{eqnarray}
\fl v_{j(1)}& = &\exp\left(\frac{1}{2}\mathrm{i}\pi\sum_{i=1}^{j-1}c^{\dagger}_{i(1)}c_{i(1)}\right),\\
\fl v_{j(2)}& = &\exp\left(\frac{1}{2}\mathrm{i}\pi\sum_{i=1}^{L}c^{\dagger}_{i(1)}c_{i(1)}\right)\exp\left(\frac{1}{2}\mathrm{i}\pi\sum_{i=1}^{j-1}c^{\dagger}_{i(2)}c_{i(2)}\right),\\
\fl v_{j(3)}& =&\exp\left(\frac{1}{2}\mathrm{i}\pi\sum_{i=1}^{L}c^{\dagger}_{i(1)}c_{i(1)}\right)\exp\left(\frac{1}{2}\mathrm{i}\pi\sum_{i=1}^{L}c^{\dagger}_{i(2)}c_{i(2)}\right)\exp\left(\frac{1}{2}\mathrm{i}\pi\sum_{i=1}^{j-1}c^{\dagger}_{i(3)}c_{i(3)}\right).
\end{eqnarray}
Above $c^{\dagger }_{j(\alpha)}$ and $c_{j(\alpha)}$ are creation and
annihilation operators with colour index $\alpha$ ($\alpha = 1,2,3$)
satisfying canonical anti-commutation relations and $n_{j(\alpha)}=c^{\dagger
}_{j(\alpha)}c_{j(\alpha)}$ is the density operator. Under such a 
transformation,
one may obtain the
Hamiltonian of a fermionic model which is equivalent to the
model (\ref{Ham-s}):
\begin{equation}
\fl H_{jj+1}= \sum _{\alpha=1}^3\;\left(c^{\dagger}_{j(\alpha)}c_{j+1(\alpha)}+
c^{\dagger}_{j+1(\alpha)}c_{j(\alpha)}\right)
\exp\left[\eta \sum _{\alpha^{'} \neq \alpha }
n_{j+\theta(\alpha^{'}-\alpha)(\alpha^{'})}\right]\label{Ham-fd}
\end{equation}
describing fermions hopping along a lattice with strong correlations
determined by 
the occupation numbers of the fermions with different colours. 
In order to apply the QISM
approach, let us now connect the fermion model (\ref{Ham-fd}) with a
Lax operator which realizes the graded Yang-Baxter relation. For this
purpose, let us first define the matrix
\begin{equation}
V_j=V_{j(1)}\otimes V_{j(2)}\otimes V_{j(3)}.
\end{equation}
Then the fermion Lax operator ${\cal L}(u)$ can be presented as
\begin{equation}
{\cal L}_{0j}(u)=V_{j+1}L_{0j}(u)V^{-1}_j.\label{Lax-f1}
\end{equation}
After lengthy algebra, we may write the fermion Lax operator in the
following way:
\begin{equation}
{\cal L}_{0j}(u)={\cal L}_{0j}^{(1)}(u){\cal L}_{0j}^{(2)}(u){\cal L}_{0j}^{(3)}(u) \label{Lax-f}
\end{equation}
where
\begin{eqnarray}
\fl {\cal L}^{(1)}_{0j}(u)& =& \left(\matrix{
g_{j(1)}^+&0&0&0&-f_1^2c_{j(1)}&0&0&0\cr
0&\tilde{g}_{j(1)}^+&0&0&0&f_1f_2c_{j(1)}&0&0\cr
0&0&\tilde{g}_{j(1)}^+&0&0&0&f_1f_2c_{j(1)}&0\cr
0&0&0&\tilde{\tilde{g}}_{j(1)}^+&0&0&0&-f_2^2c_{j(1)}\cr
\mathrm{i}f_1^2c^{\dagger}_{j(1)}&0&0&0&g^-_{j(1)}&0&0&0\cr
0&-\mathrm{i}f_1f_2c^{\dagger}_{j(1)}&0&0&0&\tilde{g}^-_{j(1)}&0&0\cr
0&0&-\mathrm{i}f_1f_2c^{\dagger}_{j(1)}&0&0&0&\tilde{g}^-_{j(1)}&0\cr
0&0&0&\mathrm{i}f_2^2c^{\dagger}_{j(1)}&0&0&0&\tilde{\tilde{g}}^-_{j(1)}
\cr }\right),\nonumber\\
\fl {\cal L}^{(2)}_{0j}(u)& =& \left(\matrix{
g_{j(2)}^+&0&-\mathrm{i}f_1^2c_{j(2)}&0&0&0&0&0\cr
0&\tilde{g}_{j(2)}^+&0&\mathrm{i}f_1f_2c_{j(2)}&0&0&0&0\cr
f_1^2c^{\dagger}_{j(2)}&0&g_{j(2)}^-&0&0&0&0&0\cr
0&-f_1f_2c^{\dagger}_{j(2)}&0&\tilde{g}_{j(2)}^-&0&0&0&0\cr
0&0&0&0&\tilde{g}_{j(2)}^+&0&-\mathrm{i}f_1f_2c_{j(2)}&0\cr
0&0&0&0&0&\tilde{\tilde{g}}_{j(2)}^+&0&\mathrm{i}f_2^2c_{j(2)}\cr
0&0&0&0&f_1f_2c^{\dagger}_{j(2)}&0&\tilde{g}_{j(2)}^-&0\cr
0&0&0&0&0&-f_2^2c^{\dagger}_{j(2)}&0&\tilde{\tilde{g}}_{j(2)}^-
\cr}\right),\nonumber\\
\fl {\cal L}^{(3)}_{0j}(u)& =& \left(\matrix{
g_{j(3)}^+&f_1^2c_{j(3)}&0&0&0&0&0&0\cr
-\mathrm{i}f_1^2c_{j(3)}^{\dagger}&g_{j(3)}^-&0&0&0&0&0&0\cr
0&0&\tilde{g}_{j(3)}^+&f_1f_2c_{j(3)}&0&0&0&0\cr
0&0&-\mathrm{i}f_1f_2c_{j(3)}^{\dagger}&\tilde{g}_{j(3)}^-&0&0&0&0\cr
0&0&0&0&\tilde{g}_{j(3)}^+&f_1f_2c_{j(3)}&0&0\cr
0&0&0&0&-\mathrm{i}f_1f_2c_{j(3)}^{\dagger}&\tilde{g}_{j(3)}^-&0&0\cr
0&0&0&0&0&0&\tilde{\tilde{g}}_{j(3)}^+&f_1^2c_{j(3)}\cr
0&0&0&0&0&0&-\mathrm{i}f_1^2c_{j(3)}^{\dagger}&\tilde{\tilde{g}}_{j(3)}^-
\cr}\right).\nonumber
\end{eqnarray}
Above we have introduced the notation:
\begin{eqnarray}
\fl 
& &g^+_{j(\alpha)}=u\exp(2\eta)+(\mathrm{i}-u\exp(2\eta))n_{j(\alpha)},~~
\tilde{g}^+_{j(\alpha)}=u\exp(\eta)+(\mathrm{i}-u\exp(\eta))n_{j(\alpha)},\nonumber\\
\fl
& &\tilde{\tilde{g}}^+_{j(\alpha)}=u+(\mathrm{i}-u)n_{j(\alpha)},~~~~~~~~~~~~~~~~~~~~~~~
g^-_{j(\alpha)}=1-(1+\mathrm{i}u\exp(2\eta))n_{j(\alpha)},\nonumber\\
\fl
& &
\tilde{g}^-_{j(\alpha)}=1-(1+\mathrm{i}u\exp(\eta))n_{j(\alpha)},~~~~~~~~~~~~
\tilde{\tilde{g}}^-_{j(\alpha)}=1-(1+\mathrm{i}u)n_{j(\alpha)},\nonumber\\
\fl
& &
f_1=\sqrt{1+u^2\exp(2\eta)},~~~~~~~~~~~~~~~~~~~~~~~~~~~
f_2=\sqrt{1+u^2}.\nonumber
\end{eqnarray}
After sophisticated algebra, indeed,  we can incorporate the fermion Lax operator (\ref{Lax-f}) into the graded Yang-Baxter relation
\begin{equation}
{\cal R}(u,v){\cal L}_{0j}(u)\otimes _s{\cal L}_{0j}(v)={\cal L}_{0j}(v)\otimes _s{\cal L}_{0j}(u){\cal R}(u,v),
\end{equation}
which is essential to the integrability of the model (\ref{Ham-fd}).
The fermion version of the quantum R-matrix comprises
\begin{equation}
{\cal R}(u,v)=W\cdot R(u,v)\cdot W^{-1},\label{R-f}
\end{equation}
where $W$ is  a $64\times 64$ diagonal matrix given by
\begin{equation}
W=\left(\begin{array}{cc}1&0\\ 0&\mathrm{i}\end{array}\right)\otimes M\otimes 
\left(\begin{array}{cc}1&0\\ 0&1\end{array}\right)
\end{equation}
with $M={\rm
diag}\left\{1,-\mathrm{i},-1,\mathrm{i},-\mathrm{i},1,-\mathrm{i},1,-1,\mathrm{i},-1,i,i,-1,-i,1\right\}$.
$R(u,v)$ stands for the quantum R-matrix for the spin model
(\ref{Ham-sd}), which is presented in \cite{zhou97}.
Above ${\otimes}_{S}$ denotes the graded tensor product
\begin{equation}
[A{\otimes }_{S}B]_{\alpha\beta, \gamma
\delta}=(-1)^{[P(\alpha)+P(\gamma)]P(\beta)}
A_{\alpha\gamma}B_{\beta\delta},
\end{equation}
with the Grassmann parities obeying the grading
$P(1)=P(4)=P(6)=P(7)=0$ and $P(2)=P(3)=P(5)=P(8)=1$.  We would like to
stress that the grading and the $W$-matrix are uniquely determined by
the Jordan-Wigner transformation (\ref{J-W-T}) and are not only an
artificial choice. However, one can see that the grading coincides
with the choice of the bosonic and fermionic degrees of freedom, i.e.,
up-spin is referred to as a bosonic degree of freedom, whereas down-spin as
a fermionic one in the basis of the auxiliary space $V(\cong C^2\otimes
C^2\otimes C^2)$ as
\begin{eqnarray}
\fl & &e_1=|\uparrow \uparrow \uparrow \rangle,~~e_2=|\uparrow \uparrow \downarrow \rangle,~~e_3=|\uparrow \downarrow \uparrow \rangle,~~e_4=|\uparrow \downarrow \downarrow \rangle,\nonumber\\
\fl & &e_5=|\downarrow \uparrow \uparrow \rangle,~~e_6=|\downarrow \uparrow \downarrow \rangle,~~e_7=|\downarrow \downarrow \uparrow \rangle,~~e_8=|\downarrow \downarrow \downarrow \rangle.\nonumber
\end{eqnarray}
It follows that 
\begin{equation}
{\cal R}(u,v){\cal T}(u)\otimes _s{\cal T}(v)={\cal T}(v)\otimes _s{\cal T}(u){\cal R}(u,v),\label{gYBA}
\end{equation}
where ${\cal T}(u)$ is the monodromy matrix 
\begin{equation}
{\cal T}(u)={\cal L}_{0L}(u)\cdots {\cal L}_{0L}(u).
\end{equation}
The graded Yang-Baxter algebra (\ref{gYBA}) ensures the commutativity of the
transfer matrix $\tau(u)={\rm Str}{\cal T}(u)$ for different spectral
parameters, i.e. $[\tau (u),\tau(v)]=0$. This implies that $\tau(u)$ can be viewed as a generating function of an infinite number of commuting conserved currents, which may be obtained through the expansion of the $\tau(u)$ in powers of $u$
\begin{equation}
\ln \tau(u)=\ln \tau(0)+H\,u+\frac{1}{2}J\,u^2+\cdots ,
\end{equation}
where $J$ is the first non-trivial conserved current next to the Hamiltonian:
\begin{eqnarray}
\fl
(-\mathrm{i})J & = &\sum_{j=1}^{L}\left\{\sum^3_{\alpha =1}(c_{j+1(\alpha )}^{\dagger}c_{j-1(\alpha )}-c_{j-1(\alpha )}^{\dagger}c_{j+1(\alpha )})\right.\nonumber\\
\fl
& &\left.\times \exp\left(\eta \sum _{\alpha^{'}\neq \alpha }n_{j+\theta(\alpha ^{'}-\alpha )(\alpha ^{'})}\right)\exp\left(\eta \sum _{\alpha^{'}\neq \alpha }n_{j-1+\theta(\alpha ^{'}-\alpha )(\alpha ^{'})}\right)\right.\nonumber\\
\fl
& &\left.-\exp\eta \, \sinh\eta \sum _{\alpha < \beta }^3\left[(c_{j-1(\alpha )}^{\dagger}c_{j(\alpha )}-c_{j(\alpha )}^{\dagger}c_{j-1(\alpha )})
(c_{j(\beta )}^{\dagger}c_{j+1(\beta )}+c_{j+1(\beta )}^{\dagger}c_{j(\beta )})\right.\right.\nonumber\\
\fl 
& &\left.\left.+(c_{j-1(\alpha )}^{\dagger}c_{j(\alpha )}+c_{j(\alpha )}^{\dagger}c_{j-1(\alpha )})
(c_{j(\beta )}^{\dagger}c_{j+1(\beta )}-c_{j+1(\beta )}^{\dagger}c_{j(\beta )})\right]\right.\nonumber\\
\fl
& &\left.\times \exp\left(\eta \sum _{\alpha^{'}\neq \alpha,\beta }n_{j-1+\theta(\alpha ^{'}-\alpha )(\alpha ^{'})}\right)\exp\left(\eta \sum _{\beta^{'}\neq \alpha,\beta }n_{j+\theta(\beta ^{'}-\beta )(\beta ^{'})}\right)\right.\nonumber\\
\fl
& &
\left.-\exp\eta \, \sinh\eta \sum _{\alpha < \beta }^3\left[(c_{j(\alpha )}^{\dagger}c_{j+1(\alpha )}+c_{j+1(\alpha )}^{\dagger}c_{j(\alpha )})
(c_{j(\beta )}^{\dagger}c_{j+1(\beta )}-c_{j+1(\beta )}^{\dagger}c_{j(\beta )})\right.\right.\nonumber\\
\fl 
& &\left.\left.+(c_{j(\alpha )}^{\dagger}c_{j+1(\alpha )}-c_{j(\alpha )}^{\dagger}c_{j-1(\alpha )})
(c_{j(\beta )}^{\dagger}c_{j+1(\beta )}+c_{j+1(\beta )}^{\dagger}c_{j(\beta )})\right]\right.\nonumber\\
\fl
& &\left.\times \exp\left(\eta \sum _{\alpha^{'}\neq \alpha,\beta }n_{j+\theta(\alpha ^{'}-\alpha )(\alpha ^{'})}\right)\exp\left(\eta \sum _{\beta^{'}\neq \alpha,\beta }n_{j+\theta(\beta ^{'}-\beta )(\beta ^{'})}\right)\right\}.
\end{eqnarray}

Therefore, we have built up an important ingredient towards the QISM approach
for the model. Next we shall discuss the integrable boundary conditions
for the fermion model with the Hamiltonian density (\ref{Ham-fd}). The
boundary conditions are known to be useful in studying
conductivity properties in such non-fermion liquids (see, for example, 
\cite{Frah,Asak}). The open boundary conditions for the spin model
(\ref{Ham-sd}) were studied in \cite{zhou3,3XY}.  Now we show that the fermion version of ${\cal R}$
satisfies the following graded reflection equations
\begin{eqnarray}
& &{\cal R}_{12}(u,v)\stackrel{1}{K_-}(u){\cal
R}_{21}(v,-u)\stackrel{2}{K_-}(v)\nonumber\\ & &=
\stackrel{2}{K_-}(v){\cal R}_{12}(u,-v)\stackrel{1}{K_-}(u){\cal
R}_{21}(-v,-u), \label{RE1}\\ 
& &{\cal R}_{21}^{{\rm St}_1{\overline {\rm
St}}_2}(v,u)\stackrel{1}{K_+^{ {\rm St}_1}}(u)\tilde{{\cal
R}}_{12}(-u,v) \stackrel{2}{K_+^{\overline {{\rm St}}_2}}(v)= \nonumber \\ &
&\stackrel{2}{K_+^{\overline {{\rm St}}_2}}(v)\tilde{{\cal R}}_{21}(-v,u)
\stackrel{1}{K_+^{{\rm St}_1}}(u) {\cal R} _{12}^{{\rm St}_1{\overline {{\rm
St}}}_2}(-u,-v), \label{RE2}
\end{eqnarray}
above  we used the conventional notation
\begin{equation}
\stackrel{1}{X} \equiv X\otimes_{S}\mathbf{I}_{V_2},\qquad 
\stackrel{2}{X} \equiv \mathbf{I}_{V_1}\otimes_{S}X,
\end{equation}
where $\mathbf{I}_{V}$ denotes the identity operator on $V$, and, as
usual, ${\cal R}_{12}={\cal P}\cdot {\cal R}$ and ${\cal R}_{21}={\cal
P}\cdot {\cal R}_{12}\cdot {\cal P}$.  Here ${\cal P}$ is the graded
permutation operator which can be represented by a $2^8\times 2^8$
matrix, i.e.,
\begin{equation}
P_{\alpha\beta, \gamma
\delta}=(-1)^{P(\alpha)P(\beta)}
\delta_{\alpha\delta}\delta_{\beta\gamma}.
\end{equation}
Furthermore, superscripts ${\rm St}_a$ and
$\overline{{\rm St}}_{a}$ denote the supertransposition in the space
with index $a$ and its inverse, respectively,
\begin{equation}
(A_{ij})^{{\rm St}}=(-1)^{[P(i)+P(j)]P(i)}A_{ji},
\qquad
(A_{ij})^{\overline{{\rm St}}} =(-1)^{[P(i)+P(j)]P(j)}A_{ji}.
\end{equation}
The graded reflection equations (\ref{RE1}) and (\ref{RE2})  together with the graded Yang-Baxter algebra (\ref{gYBA}) and the following properties
\begin{eqnarray}
{\cal R}_{12}(u,v){\cal R}_{21}(v,u) & = & 1,\\
\tilde{{\cal R}}_{21}^{\overline {{\rm St}}_1}(-v,u){\cal R}_{12}^{\overline {{\rm St}}_2}(u,-v) & = & 1,\label{tildeR1}\\
\tilde{{\cal R}}_{12}^{{\rm St}_2}(-u,v){\cal R}_{21}^{{\rm St}_1}(v,-u) & = & 1.\label{tildeR2}
\end{eqnarray}
assure that the double-row transfer matrix
\begin{equation}
\tau (u)=Str_{0}K_{+} (u){\cal T}(u)K_{-}(u){\cal T}^{-1}(-u)
\label{DTM}
\end{equation}
commutes for different spectral parameters, proving the integrability
of the model with open boundary conditions. After a lengthy
calculation, we find that the left boundary $K_-(u)$-matrix is given by
\begin{equation}
\fl
K_-^{(m)}(u) = \frac {1}{\lambda _-}
\left ( \begin {array} {cccccccc}
A_-(u)  &0&0&0&0&0&0&0\\
0&B_-(u) &0&0&0&0&0&0 \\
0&0&C_-(u) &0&0&0&0&0\\
0&0&0&D_-(u)&0&0&0&0\\
0&0&0&0&E_-(u)&0&0&0\\
0&0&0&0&0&F_-(u)&0&0\\
0&0&0&0&0&0&G_-(u)&0\\
0&0&0&0&0&0&0&H_-(u)
\end {array}  \right ),\label{Km}
\end{equation}
where for $m=1$ we have 
\begin{eqnarray}
A_-(u)&=&(c_-+u)(e^{2\eta} c_- +u)(e^{4\eta} c_-+u),\no\\
B_-(u)&=&(c_--u)(e^{2\eta} c_- +u)(e^{4\eta}c_-+u),\no\\
C_-(u)&=&(c_--u)(e^{2\eta} c_- +u)(e^{4\eta}c_-+u),\no\\
D_-(u)&=&(c_--u)(e^{2\eta} c_--u)(e^{4\eta} c_-+u),\no\\
E_-(u)&=&(c_--u)(e^{2\eta} c_- +u)(e^{4\eta}c_-+u),\no\\
F_-(u)&=&(c_--u)(e^{2\eta} c_--u)(e^{4\eta} c_-+u),\no\\
G_-(u)&=&(c_--u)(e^{2\eta} c_--u)(e^{4\eta} c_-+u),\no\\
H_-(u)&=&(c_--u)(e^{2\eta} c_- -u)(e^{4\eta} c_--u).\no\\
\lambda _-& =& \frac{1}{e^{6\eta}c_-^3}\no
\end{eqnarray}
while  $m=2$
\begin{eqnarray}
A_-(u)&=&E_-(u)=(c_-+u)(c_-+e^{2\eta}u),\no\\
B_-(u)&=&C_-(u)=F_-(u)=G_-(u)=(c_-+u)(c_--e^{2\eta}u),\no\\
D_-(u)&=&H_-(u)=(c_--u)(c_--e^{2\eta}u),\no\\
\lambda _-& =& \frac{1}{c_-^2}\no
\end{eqnarray}
and  $m=3$
\begin{eqnarray}
A_-(u)&=&C_-(u)=E_-(u)=G_-(u)=(c_-+u),\no\\
B_-(u)&=&D_-(u)=F_-(u)=H_-(u)=(c_--u),\no\\
\lambda _-& =& \frac{1}{c_-}.\no
\end{eqnarray}
These results coincide with those obtained  for the spin model studied 
in ref.~\cite{3XY}. 
However, its companion, the right boundary
$K_+(u)$-matrix:
\begin{equation}
\fl K_+^{(l)}(u)= \left ( \begin {array} {cccccccc} A_+(u)
&0&0&0&0&0&0&0\\ 0&B_+(u) &0&0&0&0&0&0 \\ 0&0&C_+(u) &0&0&0&0&0\\
0&0&0&D_+(u)&0&0&0&0\\ 0&0&0&0&E_+(u)&0&0&0\\ 0&0&0&0&0&F_+(u)&0&0\\
0&0&0&0&0&0&G_+(u)&0\\ 0&0&0&0&0&0&0&H_+(u)
\end {array}  \right ) ,   \label{Kp}
\end{equation}
for $l=1$ 
\begin{eqnarray}
A_+(u)&=&(e^{6\eta}c_+u-1)(e^{4\eta}c_+u -1)
(e^{2\eta}c_+u-1) ,\no\\
B_+(u)&=&-e^{4\eta}(e^{2\eta}c_+u+1)(e^{4\eta}c_+u -1)
(e^{2\eta}c_+u-1) ,\no\\
C_+(u)&=&-e^{2\eta}(e^{2\eta}c_+u+1)(e^{4\eta}c_+u -1)
(e^{2\eta}c_+u-1),\no\\ D_+(u)&=&e^{6\eta}(e^{2\eta}c_+u+1)(c_+u +1)
(e^{2\eta}c_+u-1),\no\\ E_+(u)&=&-(e^{2\eta}c_+u+1)(e^{4\eta}c_+u -1)
(e^{2\eta}c_+u-1) ,\no\\ F_+(u)&=&e^{4\eta}(c_+u+1)(e^{2\eta}c_+u +1)
(e^{2\eta}c_+u-1) ,\no\\ G_+(u)&=&e^{2\eta}(c_+u+1)(e^{2\eta}c_+u +1)
(e^{2\eta}c_+u-1),\no\\ H_+(u)&=&-e^{4\eta}(c_+u+e^{2\eta})(c_+u +1)
(e^{2\eta}c_+u+1).\no 
\end{eqnarray}
 for $l=2$ 
\begin{eqnarray}
A_+(u)&=&-B_+(u)=(e^{6\eta}c_+u-1)(e^{4\eta}c_+u -1),\no\\
C_+(u)&=&-D_+(u)=-e^{2\eta}(e^{2\eta}c_+u+1)(e^{4\eta}c_+u -1),\no\\
E_+(u)&=&-F_+(u)=-(e^{2\eta}c_+u+1)(e^{4\eta}c_+u -1),\no\\
G_+(u)&=&-H_+(u)=e^{2\eta}(c_+u+1)(e^{2\eta}c_+u +1),\no 
\end{eqnarray}
 for $l=3$
\begin{eqnarray}
 A_+(u)&=&-B_+(u)=e^{2\eta}(e^{4\eta}c_+u -1),\no\\
C_+(u)&=&-D_+(u)=-(e^{4\eta}c_+u -1),\no\\
E_+(u)&=&-F_+(u)=-e^{2\eta}(c_+u +1),\no\\ G_+(u)&=&-H_+(u)=(c_+u+1),\no
\end{eqnarray}
is  different from that in the non-graded case. Thus
the graded reflection equations (\ref{RE1}) and (\ref{RE2}) warrant
the following boundary terms to be integrable:
\begin{eqnarray}
\fl B_{1}^{(m)} &=&
\left\{\begin{array}{ll}
\frac{1}{c_- \exp(2\eta)}\left[\begin{array}{l}\exp(-2\eta)\sum _{\alpha=1}^{3}n_{1(\alpha)} \\
+2\exp(-\eta)\sinh \eta \sum ^{3}_{\mbox{\scriptsize $\begin{array}{l}\alpha,\beta=1\\ \alpha \neq 
\beta \end{array}$}}n_{1(\alpha)} n_{1(\beta)}\\
+4\sinh^2 \eta\; n_{1(1)}
	  n_{1(2)} n_{1(3)}\end{array}\right], & \rm{for}~~m=1\\
\frac{\exp(\eta)}{c_-}\left[\exp(-\eta) \sum _{\alpha=2}^{3}n_{1(\alpha)} 
          +2\sinh \eta n_{1(2)} n_{1(3)}\right], & 
\rm{for}~~m=2\\
\frac{1}{c_-}n_{1(3)}, &\rm{for}~~m=3
\end{array}\right.\label{Ham-bt-1}\\
\fl 
B_{L}^{(l)} &=&
\left\{\begin{array}{ll}
\frac{1}{c_+}\left[\begin{array}{l}\exp(-2\eta) 
	  \sum _{\alpha=1}^{3}n_{L(\alpha)} \\
+2\exp(-\eta)\sinh \eta \sum ^{3}_{\mbox{\scriptsize $\begin{array}{l}\alpha,\beta=1\\ \alpha \neq 
\beta \end{array}$}}n_{L(\alpha)} n_{L(\beta)}\\
+4\sinh^2 \eta\; n_{L(1)}
	  n_{1(2)} n_{L(3)}\end{array}\right], & \rm{for}~~l=1\\
\frac{1}{c_+exp(\eta)}\left[\exp(-\eta) 
	  \sum _{\alpha=1}^{2}n_{L(\alpha)} 
          +2\sinh \eta n_{L(1)} n_{L(2)}\right], & 
\rm{for}~~l=2\\
\frac{1}{c_+}n_{L(1)}, &\rm{for}~~l=3
\end{array}\right.\label{Ham-bt-2}
\end{eqnarray}
where $c_{\pm}$ are parameters describing boundary effects. With
the different choices of the pair $(m,l)~~m,l=1,2,3$, 
there exist nine classes of integrable boundary terms compatible with
the integrability of the model (\ref{Ham-fd}).

 So far, we have performed the fermionization of the one-dimensional
Bariev model of three coupled XY chains. By verifying the graded
Yang-Baxter relation, the fermion Lax operator and the quantum R-matrix
are derived  explicitly. Further, the integrable boundary conditions for
the fermion model are discussed. Our results provide a the starting point 
towards the algebraic Bethe ansatz for the model with both periodic
and open boundary conditions by means of the QISM.

{\bf Acknowledgments} 
 A.F. and X.W.G. thank CNPq (Conselho Nacional de Desenvolvimento
Cient\'{\i}fico e Tecnol\'ogico) and FAPERGS (Funda\c{c}\~{a}o de
Amparo \~{a} Pesquisa do Estado do Rio Grande do Sul) for financial
support.  X.W.G. gratefully acknowledges
the hospitality of the Institut f\"{u}r Physik, Technische
Universit\"{a}t Chemnitz.  J.L. and  H.Q.Z. acknowledges
the support from the Australian Research Council.

\end{document}